\documentclass{emulateapj}

\begin{document}

\title{Phosphorus enrichment  by  ONe novae in the Galaxy}

\author{Kenji Bekki} 
\affil{
ICRAR,
M468,
The University of Western Australia
35 Stirling Highway, Crawley
Western Australia, 6009, Australia
}

\author{Takuji Tsujimoto}
\affil{
National Astronomical Observatory of Japan, Mitaka-shi, Tokyo 181-8588, Japan}

\begin{abstract}
Recent observations have shown that [P/Fe] in the Galactic stars decreases with
increasing [Fe/H] for [Fe/H]$\gtrsim -1$ whereas it is almost subsolar for [Fe/H]$\lesssim -2$. 
These [P/Fe] trends with [Fe/H] have not been well reproduced by previous theoretical
models incorporating phosphorus (P) enrichment only by core collapse supernoave.
We here show, for the first time, that the trends can be naturally explained by
our new models incorporating P enrichment by oxygen-neon (ONe) novae
which occur at the surface of massive
white dwarfs whose masses are larger than $1.25 M_{\odot}$
with a metallicity-dependence rate.
We also show that the observations can be 
better reproduced by the models by assuming that
(i) the total mass of gaseous  ejecta per ONe nov ($M_{\rm ej}$) 
is as high as $4 \times 10^{-5} M_{\odot}$
and (ii)  the number of such novae per unit mass ($N_{\rm ONe}$) is as large as 0.01 at 
[Fe/H]$\approx -3$.
The assumed $M_{\rm ej}$ is consistent with observations,
and the high $N_{\rm ONe}$ is expected from recent theoretical models for ONe nova
fractions.
We predict that (i) [P/Fe] increases with
increasing [Fe/H] for $-2 \lesssim {\rm [Fe/H]} \lesssim -1$ and (ii) [P/Fe] and [Cl/Fe] trends with [Fe/H] 
are very similar   with each other due to very large yields of P and Cl
from  ONe nova.
It is thus  worthwhile for future observations to 
assess the validity of the proposed P enrichment by ONe novae
by confirming or ruling out these two predictions.

\end{abstract}

\section{Introduction}

Recent spectroscopic observations of the Galactic stars have revealed intriguing
abundance patterns of  phosphorus (P) depending on metallicities 
(e.g., Caffau et al. 2011; Roederer et al. 2014; Hinkel et al. 2020; 
Masseron et al. 2020, M20; Snedin et al. 2021; Maas et al. 2022; Nandakumar et al. 2022).
For example, [P/Fe] is observed
to decrease with increasing [Fe/H] at $-1 \lesssim   {\rm [Fe/H]} \lesssim 0.3$
(e.g., Caffau et al. 2011; Nandakumar et al. 2022), which is similar to
[$\alpha$/Fe] (e.g., [Mg/Fe]) trends with [Fe/H].
Roederer et al. (2014) found that [P/Fe] at $-4 \lesssim {\rm [Fe/H]} \lesssim -2$ is
either slightly above solar or subsolar with no stars having higher ($>0.3$) [P/Fe],
which is in a striking contrast with the observed trend of [$\alpha$/Fe] with [Fe/H].
These characteristic trends of [P/Fe] with [Fe/H], which need to be reproduced
by any theory of galaxy formation,  are summarized in
Figure 1.

Cescutti et al. (2012, C12) first tried to reproduce the observed trend of [P/Fe] with [Fe/H]
from Caffau et al. (2011). 
They demonstrated that their models with the adopted P yields
of core collapse supernovae (CCSNe) 
being by a factor of $3$ higher than those 
from Kobayashi et al. (2006, K06)
can reproduce well the decreasing [P/Fe] with increasing [Fe/H] at [Fe/H]$\gtrsim -1$.
However,
it is unclear whether such an ad hoc yield increase is
really reasonable and realistic  in the nucleosynthesis of CCSNe.
Furthermore the observed sub-solar [P/Fe] at [Fe/H]$\lesssim -2$ (e.g., Roederer et al. 2014, 2016)
is inconsistent with the models by C12
which predicted high [P/Fe] ($\gtrsim 0.3$) at [Fe/H]$\lesssim -2$. 
Other chemical evolution  models  with P enrichment by CCSNe and rotating massive stars
by Prantzos et al. (2018) predicted an almost constant [P/Fe] ($\sim$ solar value)
and accordingly cannot reproduce the observed [P/Fe] trend with [Fe/H] either.
Therefore, polluter other than  CCSNe (e.g., AGB stars and novae) are required to reproduce
the [P/Fe] evolution of the Galaxy.
As shown in M20, the P yields of AGB stars are significantly lower than those of novae, which suggests
that P-enrichment by novae would need to be first considered in the Galactic chemical evolution models.

Previous and recent nucleosynthesis models for novae formed from
oxygen-neon-magnesium (ONe) white dwarfs (WDs) have shown that
the mass fractions of P in  ONe novae
with WD masses ($M_{\rm WD}$)  being
1.25 and $1.35M_{\odot}$
(referred to as ``P-rich'' ONe nova just for convenience)
are as high as $0.01$  corresponding
to the mass fractions
by a factor of $\approx 1000$ larger than those of CCSNe in some of the models
(e.g., Jos\'e \& Hernanz 1998, JH98; Starrfield et al. 1998, S98).
This suggests that P enrichment by these ONe novae can be crucial in
[P/Fe] evolution even if the total ejecta mass is relatively
small compared to those from CCSNe: chemical enrichment by P-rich ONe nova should be 
more seriously considered in chemical evolution models
of the Galaxy. However, no chemical evolution  models have ever investigated 
whether P enrichment by ONe novae can explain the observed characteristic [P/Fe] trends
with [Fe/H] in the Galaxy.

The purpose of this {\it Letter} is thus  to demonstrate,
for the first time, that P enrichment by ONe nova 
is responsible for the observed unique [P/Fe] trend with [Fe/H] in the Galactic stars.
In this investigation, we particularly consider a recent theoretical prediction
from Kemp et al. (2022, K22) that nova rates strongly depend on metallicities ($Z$)
 such that
they are significantly lower at higher $Z$.
As illustrated in Figure 1,
the combination of high P yields from P-rich ONe novae and $Z$-dependent
nova rates can reproduce
the observed unique [P/Fe]$-$[Fe/H] relation of the Galactic stars in
the present study.
Using one-zone chemical evolution models incorporating P enrichment by ONe novae
and the above theoretical prediction of P yields from ONe novae,
we investigate how [P/Fe] evolution with [Fe/H] depends on the model parameters
such as the total ejecta mass per ONe nova ($M_{\rm ej}$) and 
the number fraction of P-rich ONe novae among all nova populations.
Our future chemodynamical models of the Galaxy incorporating
P enrichment  will predict
how the  [P/Fe]$-$[Fe/H] relation is different in different regions of the Galaxy.

\begin{figure}
\plotone{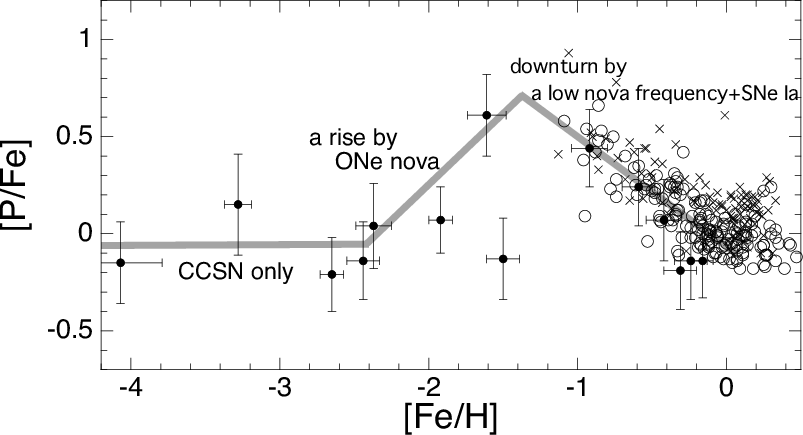}
\figcaption{
Distribution of the Galactic P-normal stars with [P/Fe]$<1$
on the [Fe/H]$-$[P/Fe] plane and illustration
of the predicted evolutionary path of the ``ONe nova'' scenario (wide gray line).
Observational results for P-normal stars
([P/Fe]$<1$)  from (i) Roederer et al. (2014, 2016),
(ii) Maas et al. (2022),
and (iii) Caffau et al. (2016, 2019), M20,
Maas et al. (2022),  and Nandakumar et al. (2022)
are plotted by
filled circles,
open circles,
and   crosses, respectively.
The gray line just illustrates how [P/Fe] evolves with [Fe/H]: the exact details
of the evolutionary paths (e.g., [Fe/H] at the maximum [P/Fe]) from
the present chemical evolution model can be different from this line.
Characteristic features  of the models are explained with short phrases 
at appropriate places
(e.g., ``CCSN only'' at [Fe/H]$<-2.4$).
\label{fig-1}}
\end{figure}

\section{The model}

We use our original code for one-zone chemical evolution of galaxies  developed in our previous studies
(Bekki \& Tsujimoto 2012, 2023) to investigate the time evolution
of [P/Fe] and [Cl/Fe] of the Galaxy.
In the present-study,
we newly incorporate (i) P enrichment by
ONe novae (JH98) and  (ii)
P yields of CCSNe that can depend on metallicities in the code (K06).
In investigating P enrichment by novae, 
we consider contributions only from P-rich ONe novae with
their accreting white dwarf masses ($M_{\rm WD}$) 
being as large as or larger than $1.25 M_{\odot}$,
because the mass fraction of P in the ejecta ($f_{\rm P}$) can be $\approx 0.01$:
Other novae from CO WDs and ONe novae with $M_{\rm WD} < 1.25 M_{\odot}$
have very low  $f_{\rm P}$ ($\ll  0.01$). 
We consider that (i) the number of novae per unit (star-forming) mass ($N_{\rm N}$),
(ii) the number of outbursts per nova ($N_{\rm burst}$), (iii) the number fraction
of P-rich ONe novae among all nova populations ($f_{\rm ONe}$),
and (iv) the total mass of ejecta per nova ($M_{\rm ej}$) are key parameters
in P evolution of the Galaxy. 

We adopt $N_{\rm burst}=10^4$ that was adopted by chemical evolution models
of the Galaxy (e.g.,  Matteucci et al. 2003; Romano et al. 2003; C12) 
and is consistent with observations (e.g., Della Valle and Izzo 2020 for a recent
review).
Since previous models adopted $N_{\rm N}=0.03$ to explain the observed lithium (Li)
abundance patterns of the Galactic stars and the present-day total nova rate 
(e.g., Cescutti et al. 2019), we consider the same value too: it should be noted here, however,
that $N_{\rm N} \approx 0.014$ at $Z=0.0001$ is predicted by  K22.
The number of P-rich ONe novae per unit mass ($N_{\rm ONe}$) is the most important
parameter of the present study  and denoted as follows;
\begin{equation}
N_{\rm ONe}=f_{\rm ONe}N_{\rm N}.
\end{equation}
K22 predicted that (i) the number fraction of ONe novae among all (CO and ONe)
novae
is about 70\%
in the Galaxy (see their Fig. 12)
and  (ii) the number fraction of P-rich ONe novae with $M_{\rm WD}>1.25 M_{\odot}$
among ONe novae is also high (roughly 0.6 in their Fig. 13). We therefore consider that
$f_{\rm ONe}$ (the above (i) multiplied by (ii)) can be as high as 
$\approx 0.4$ 
corresponding to  $N_{\rm ONe}=0.012$
for $N_{\rm N}=0.03$.

K22 calculated the time delay between 
star formation and the onset of the  nova outburst for each binary star system
and thereby derived the delay time ($t_{\rm delay}$) 
distribution for ONe novae. K22 showed that
a very strong peak is at $t_{\rm delay}=500$ Myr both for $Z=0.0001$ and 0.03
and the majority of the novae have $t_{\rm delay} <2$ Gyr (see their Fig. 11).
Given that the novae recurrence time for $M_{\rm WD}>1.2 M_{\odot}$ is
shorter than $2.8 \times 10^4$ yr  (Truran \& Livio 1986), 
$N_{\rm burst}=10^4$ means that all outbursts for P-rich ONe nova can occur
within a timescale of less than $ 10^9$ yr after star formation.
Considering these previous results,
we assume that the first and last  outbursts occur  $10^8$ and $10^9$ yr after
star formation, respectively, for P-rich ONe novae.

A key element of the present model is that 
$N_{\rm ONe}$ depends on metallicities ([Fe/H]) as follows:
\begin{equation}
N_{\rm ONe}=N_{\rm ONe, 0}+\alpha_{\rm Z} \times ({\rm  [Fe/H]-[Fe/H]_0)},
\end{equation}
where $N_{\rm ONe, 0}$ is $N_{\rm ONe}$ at the adopted initial [Fe/H]
(i.e., ${\rm [Fe/H]_0}=-3$)
and $\alpha_{\rm Z}$ determines the slope of the 
$N_{\rm ONe}-{\rm  [Fe/H]}$ relation.
Since K22 predicts that $N_{\rm N}$ is significantly higher for lower $Z$ (thus lower
[Fe/H]),  $\alpha_{\rm Z}$ should be negative: $N_{\rm ONe}$ should be larger
for lower [Fe/H] too. We mainly show the results of the models with 
$N_{\rm ONe, 0}=0.01$ and
$\alpha_{\rm Z}=-3.0 \times 10^{-3}$, because they can better reproduce observations.
We investigate the models with $M_{\rm ej}$ ranging from $10^{-5} M_{\odot}$ to
$ 6 \times 10^{-5} M_{\odot}$,
which is consistent with observations (e.g., Dell Valle \& Izzo 2020).
Since the mass fractions of Cl ($f_{\rm Cl}$) in gaseous ejecta
of P-rich ONe novae are very high too ($>10^{-3}$; JH98), it is expected
that [P/Fe] and [Cl/Fe] evolutionary trends  with [Fe/H] are  very similar with each other.
We adopt $f_{\rm P}=0.01$ and $f_{\rm cl}=0.004$ from JH98 for all models,
though these values can be even higher (e.g., S98).

We consider that the mass fraction of P in the ejecta of CCSNe  ($f_{\rm P, CCSN}$,
calculated from tables in K06)
for a given initial mass function of stars (IMF) is
also fundamentally important  for [P/Fe] evolution.
Given that P yields of CCSNe depend on metallicities (K06),
we assume that $f_{\rm P, CCSN}$
can depend on metallicities ($Z$) as follows:
$f_{\rm P, CCSN}=1.1 \times 10^{-5}+\beta_{\rm Z} \times Z$,
where $\beta_{\rm Z}$ determines the slope of the metallicity-dependence.
We use this simplified model for $f_{\rm P, CCSN}$, because
P yields are given only for four  different $Z$ in K06.
Given a possible uncertainty in the metallicity-dependence,
we  investigate
the models with $\beta_{\rm Z}=0$ 
(no metallicity-dependence), 0.001,  0.002, and 0.004.
We choose 
$1.1 \times 10^{-5}$ at $Z=0$,
because the present models can reproduce the observed [P/Fe] (sub-solar values)  at 
$-3\lesssim {\rm [Fe/H]} \lesssim  -2$ quite well.
K06 predicted constantly low [Cl/Fe] ($\approx -0.4$) over $-4 \le {\rm [Fe/H]} \le 0$
due to the low mass fraction of Cl in the ejecta of CCSNe ($f_{\rm Cl, CCSN}$),
even though  K06 included chemical enrichment by hypernovae with rather large Cl yields:
Chemical evolution  models without hypernovae can  show even lower [Cl/Fe].
Since such low [Cl/Fe] is inconsistent with observations
(Maas \& Pilachowski 2021), we investigate models with  a wider range of
$f_{\rm Cl, CCSN}$.

\begin{figure}
\epsscale{1.3}
\plotone{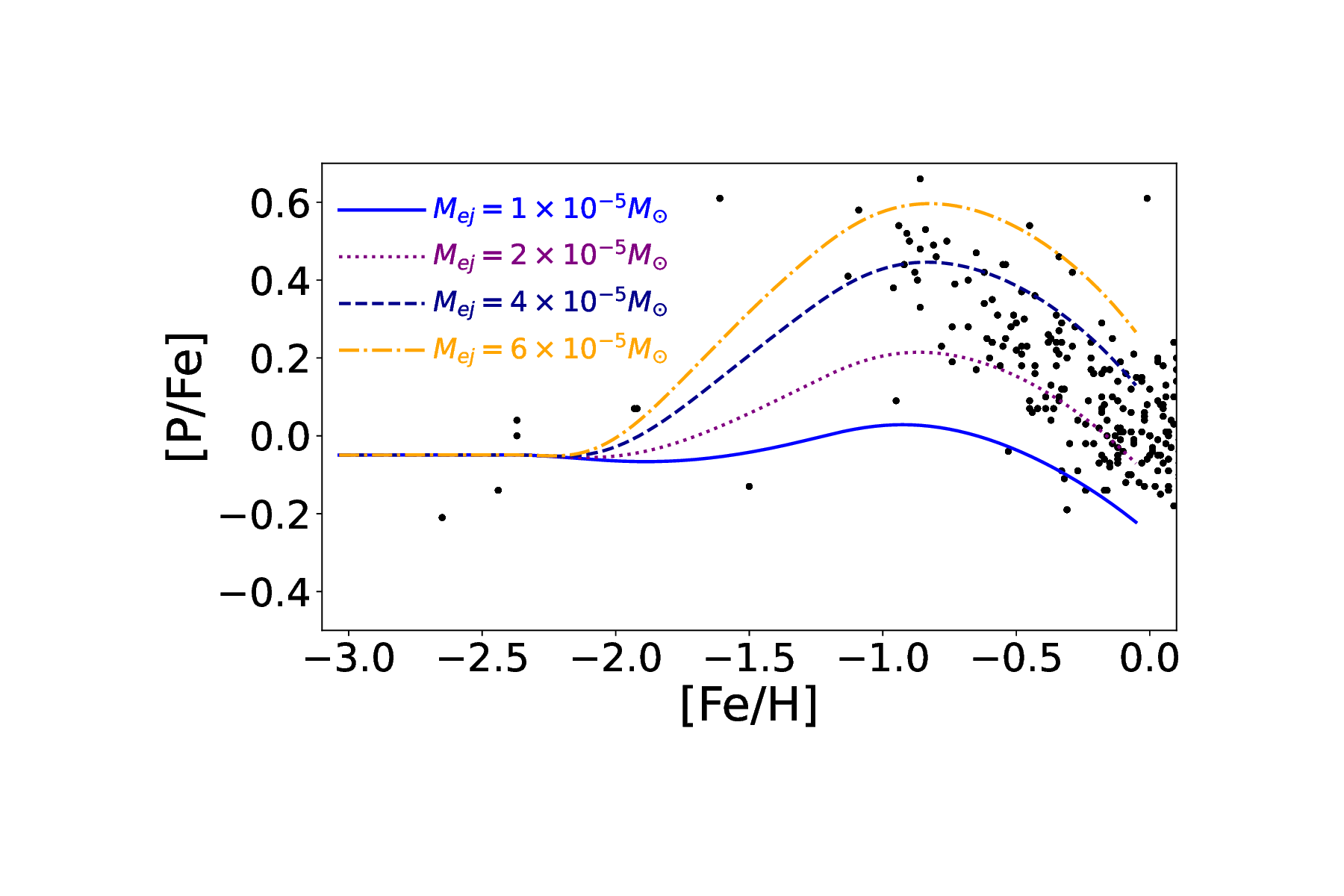}
\figcaption{
Evolutionary paths of  four models with 
$M_{\rm ej}=1 \times 10^{-4} M_{\odot}$ (blue solid),
$2 \times 10^{-4} M_{\odot}$ (purple dotted),
$4 \times 10^{-4} M_{\odot}$ (dark blue dashed, fiducial model),
and $6 \times 10^{-4} M_{\odot}$ (orange dot-dashed)
on the [Fe/H]-[P/Fe] plane.
These models adopt (i) [Fe/H]-dependent $N_{\rm ONe}$ ($\alpha_{\rm Z}=3 \times 10^{-3}$),
(ii) $N_{\rm ONe,0}=0.01$, and (iii) fixed mass fraction of P in ejecta of CCSNe
for the adopted Salpeter IMF
($f_{\rm P, CCSN}=1.1 \times 10^{-5}$ and
$\beta_{\rm Z}=0$).
Small black dots are observational data  taken from Figure 1.
\label{fig-2}}
\end{figure}

We adopt the Salpeter IMF of stars with the slope ($\alpha$) of $-2.35$,
the lower and upper  mass cut-off being $0.1 M_{\odot}$ and $50 M_{\odot}$, respectively.
The gas infall timescale and
the the number of SNe Ia per unit mass
are  set to be 3 Gyr and 0.085, respectively. The star formation rate (SFR)
at each time step is assumed to
be proportional to the gas mass ($M_{\rm g}$), i.e.,  SFR$=C_{\rm sf} M_{\rm g}$,
where $C_{\rm sf}$ is chosen such that the final [Fe/H] is close to the solar value.
We investigate 13 Gyr chemical  evolution of the Galaxy for
the initial gaseous [Fe/H] of $-3$.
The model with $N_{\rm ONe, 0}=0.01$, $\alpha_{\rm Z}=-3 \times 10^{-3}$,
$f_{\rm P}$=0.01,  $f_{\rm P, CCSN}=1.1 \times 10^{-5}$ at [Fe/H]$=-3$, 
$\beta_{\rm Z}=0$, and $M_{\rm ej}=4 \times 10^{-5} M_{\odot}$
is selected and referred to as the ``fiducial''  model,
because it can best explain both  the observed peak [P/Fe]
at [Fe/H]$\approx -0.8$ and [P/Fe]$-$[Fe/H] relation among all models investigated.

\begin{figure}
\epsscale{1.3}
\plotone{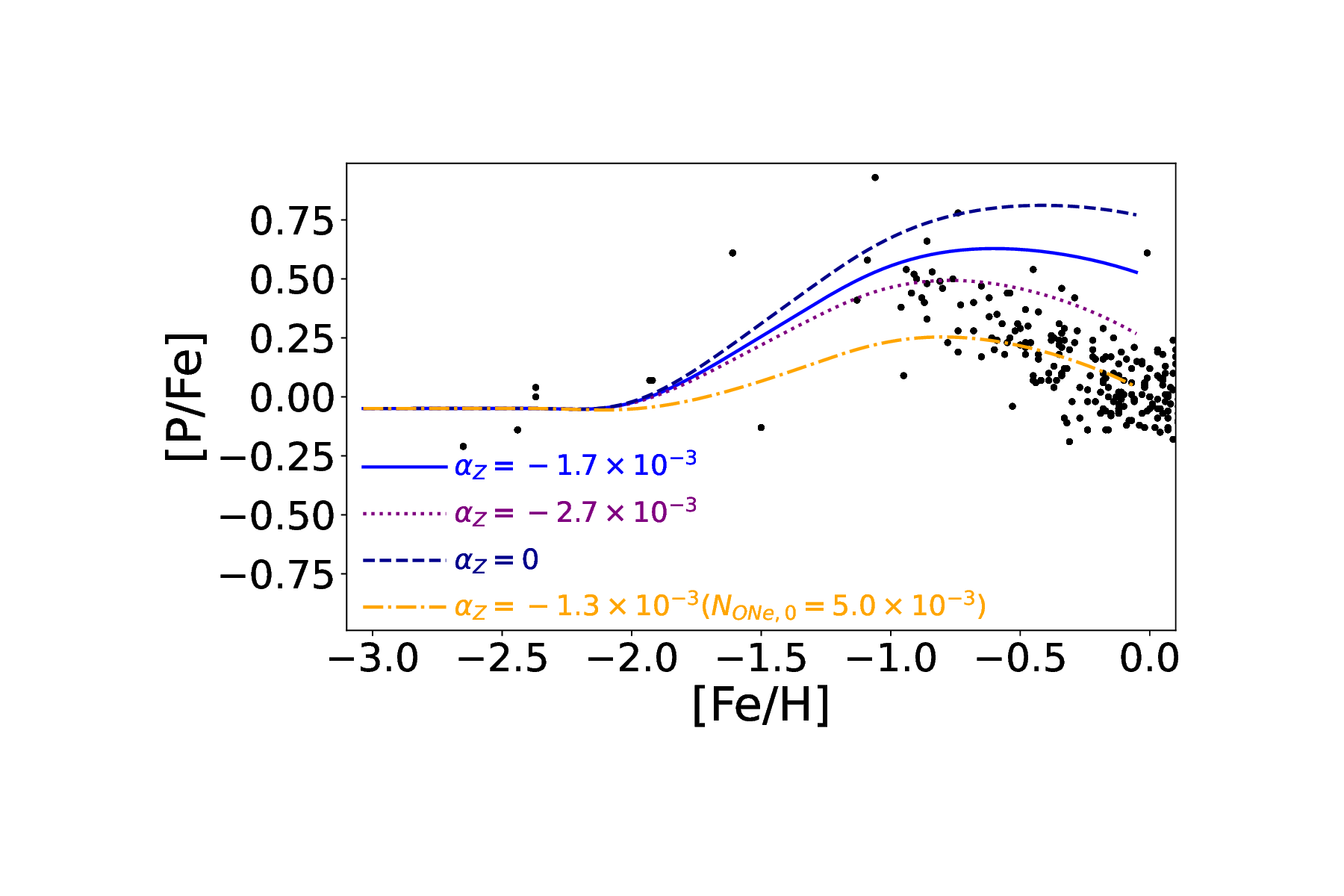}
\figcaption{
Same as Figure 2 but for four models with
$\alpha_{\rm Z}=-1.7 \times 10^{-3}$ and $N_{\rm ONe,0}=0.01$ (blue solid),
$\alpha_{\rm Z}=-2.7 \times 10^{-3}$ and $N_{\rm ONe,0}=0.01$,
(purple dotted),
$\alpha_{\rm Z}=0$ and $N_{\rm ONe,0}=0.01$,
(dark blue dashed),
and $\alpha_{\rm Z}=-1.3 \times 10^{-3}$ and $N_{\rm ONe,0}=0.003$,
(orange dot-dashed).
Only the last model has a different $N_{\rm ONe, 0}$ from other three in this figure
and those in Figure 2.
Other model parameters in the four are exactly the same as those in the fiducial model.
\label{fig-3}}
\end{figure}

\section{Results}

Figure 2 describes how the evolutionary path of the Galaxy on the [Fe/H]-[P/Fe] plane
depends on $M_{\rm ej}$ in four models with the same $\alpha_{\rm Z}$ ($=3 \times 10^{-3}$)
yet different $M_{\rm ej}$.
Clearly, all models show (i) constantly low [P/Fe] at [Fe/H] $\lesssim -2$,
(ii) increasing [P/Fe] with increasing [Fe/H] at
$-2 \lesssim  {\rm [Fe/H]} \lesssim  -0.8$,
and (iii) decreasing [P/Fe] with increasing [Fe/H] at [Fe/H]$ \gtrsim -0.8$,
though the peak [P/Fe] at [Fe/H]$\approx -0.8$ strongly depends on $M_{\rm ej}$.
The above result (ii) is due to the earlier enrichment by P-rich ONe novae
that can occur only $10^8$ yr after star formation.
The result (iii) is caused by a significant decrease of [P/Fe] due to
the combination effect of  lower ONe nova rates at higher [Fe/H] and efficient
iron production by SNe Ia.
This result implies that the observed [Fe/H]$-$[P/Fe] relation can
constrain $M_{\rm ej}$: If $N_{\rm ONe, 0} \le  0.01$,
then $M_{\rm ej} \le 10^{-5} M_{\odot}$ can be ruled out
in the ONe nova scenario.
Given that different $M_{\rm ej}$ is expected among ONe novae with different WD masses (JH98),
Figure 2 also suggests that the observed [P/Fe] dispersion at a given [Fe/H]
could be due to the different degrees of P enrichment in ISM polluted by different
ONe novae.

Figure 3 shows that the models with no or weaker dependence of $N_{\rm ONe}$ on
[Fe/H] ($-2.7 \le \alpha_{\rm Z} \le 0$) cannot reproduce the observed
[Fe/H]$-$[P/Fe] relation at [Fe/H]$\gtrsim -1$ so well compared
with the fiducial  model with
$\alpha_{\rm Z} = -3 \times 10^{-3}$.
These results clearly demonstrate that steep metallicity-dependent nova rates
are essential to explain the observed characteristic [Fe/H]$-$[P/Fe] relation.
Although the model with a lower initial number
of P-rich ONe novae  ($N_{\rm ONe,0}=5 \times 10^{-3}$)
and a shallower metallicity-dependent nova rate
($\alpha_{\rm Z} = -1.3 \times 10^{-3}$)
can reproduce 
[Fe/H]$-$[P/Fe] relation at [Fe/H]$\gtrsim -0.6$ ,
the peak [P/Fe] at [Fe/H]$\approx -0.8$ is significantly lower than the observed [P/Fe].
This result suggests that $N_{\rm ONe,0}$  needs to be higher ($>5 \times 10^{-3}$)
for  $M_{\rm ej} \approx  4 \times  10^{-5} M_{\odot}$ 
to explain the observed high [P/Fe] around [Fe/H]=$-0.8$,

It is clear from Figure 4 that four models with different metallicity-dependent P yields of CCSNe
and without ONe novae all fail to explain the observed characteristic [Fe/H]$-$[P/Fe] relation.
This demonstrates that chemical enrichment by P-rich ONe novae is essential
to reproduce  the characteristic relation.
In order to reproduce the observed [P/Fe] decrease with increasing [Fe/H] at [Fe/H]$\gtrsim -0.8$
the models without ONe novae  need to assume that $f_{\rm P, CCSN}$ is  smaller for
higher [Fe/H] at [Fe/H]$\gtrsim -0.8$. 
Such a contrived metallicity-dependent $f_{\rm P, CCSN}$  appears to be
unphysical, given that P yields are larger for higher metallicities in CCSNe with
different initial stellar masses (K06).
We therefore conclude that CCSNe alone cannot be responsible for the observed
unique [Fe/H]$-$[P/Fe] relation in the Galaxy.

As shown in Figure 5,
irrespectively of $f_{\rm Cl, CCSN}$,
the present models can reproduce the observed [Cl/Fe] at ${\rm [Fe/H]} \gtrsim -0.5$
pretty well.
As P-rich ONe novae start to enrich the ISM around
[Fe/H]$= -2$, [Cl/Fe] increases rapidly with increasing [Fe/H] to reach  its peak
at [Fe/H]$\approx -0.8$. [Cl/Fe] subsequently decreases with increasing [Fe/H] due to
the chemical enrichment by SNe Ia  and the lower rates of P-rich and Cl-rich  ONe novae  at
higher metallicities.  The derived [Cl/Fe] evolution at [Fe/H]$\gtrsim -0.5$ is qualitatively similar to
the observed one, though the observed dispersion of [Cl/Fe] at a given [Fe/H] appears to be large.
These behaviors of [Cl/Fe]  are very similar to those of [P/Fe], which
is a unique prediction of the ONe nova scenario to be tested against future observations.
Since there is no observational data point at [Fe/H]$\lesssim -0.5$, it is not clear whether
the characteristic evolutionary path of the ONe nova scenario
on the [Fe/H]$-$[Cl/Fe] plane can be consistent with
the corresponding observation.
Figure 5 also demonstrates that peak
[Cl/Fe] at [Fe/H]$\approx  -0.8$ can be only slightly different between
models with different $f_{\rm Cl, CCSN}$,
even if the adopted initial [Cl/Fe] values are quite different.

\begin{figure}
\epsscale{1.3}
\plotone{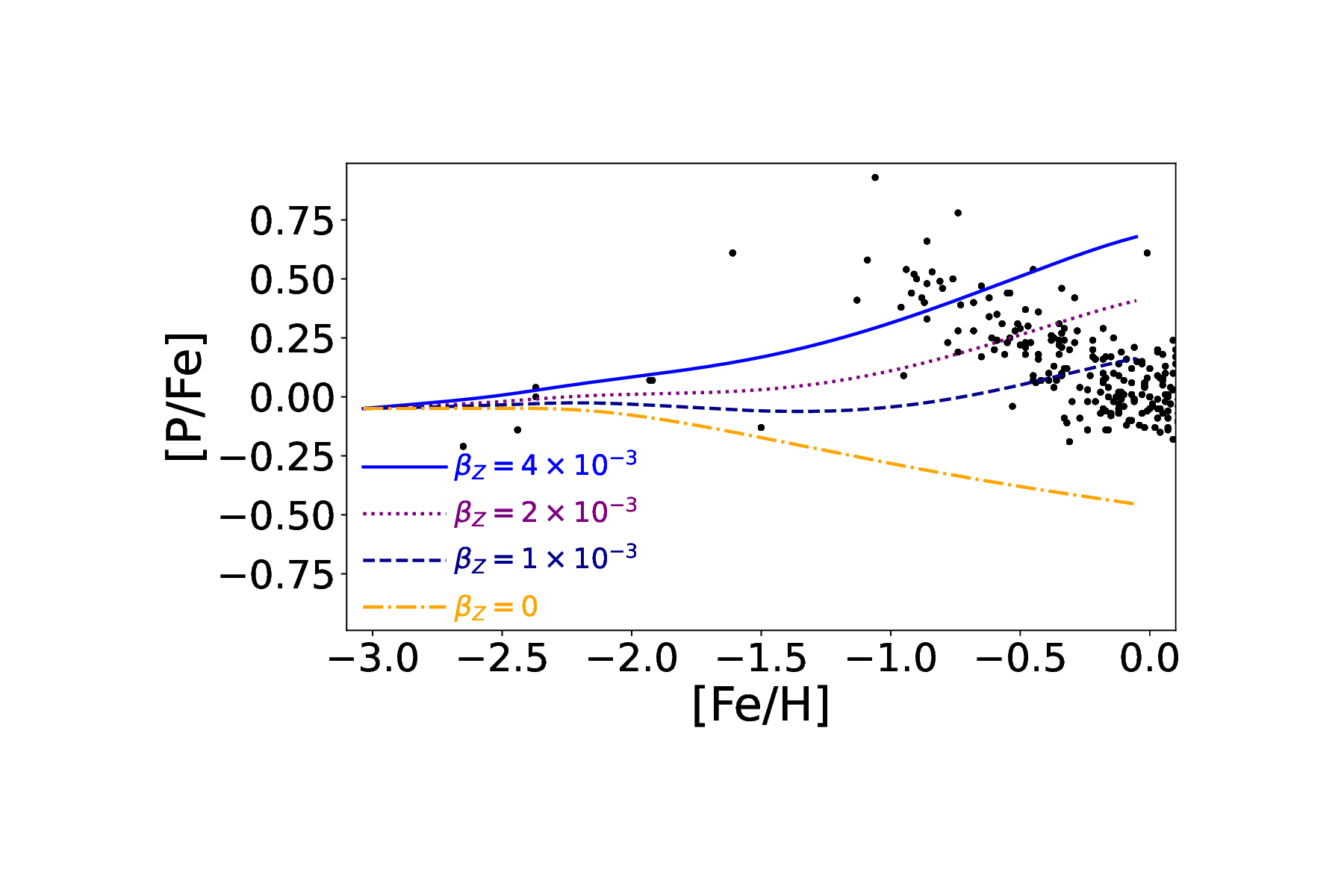}
\figcaption{
Same as Figure 2 but for four models with
different slopes of metallicity-dependent P yields of CCSNe, i.e.,
$\beta_{\rm Z}=4 \times 10^{-3}$  (blue solid),
$\beta_{\rm Z}=2 \times 10^{-3}$,
(purple dotted),
$\beta_{\rm Z}=1 \times 10^{-3}$,
(dark blue dashed),
and $\beta_{\rm Z}=0$
(orange dot-dashed, no metallicity-dependence of P yields).
\label{fig-4}}
\end{figure}

\section{Discussion}
The required $N_{\rm ONe,0} \approx 0.01$ at [Fe/H]$=-3$ 
appears to be quite high for $M_{\rm ej}=4 \times 10^{-5}$,
though such a high fraction of  ONe novae with $M_{\rm WD} > 1.2M_{\odot}$
is expected in the Galactic thin disk (see Fig. 13 in K22).
Recent observational studies of WDs by the Gaia space mission
have found that the  masses of progenitor stars that lead to  WDs with $M_{\rm WD} \gtrsim 1.2 M_{\odot}$
is around $7 M_{\odot}$ (Tremblay et al. 2024). 
 Accordingly, the number of stars
with their initial masses ($M$) ranging from $7M_{\odot}$ to $10 M_{\odot}$ 
(corresponding to $M_{\rm WD}=1.35 M_{\odot}$; Truran and Livio 1986)
per unit mass 
is $3.6 \times 10^{-3}$ 
for $\alpha=2.35$ (the Salpeter IMF) 
and $9.1 \times 10^{-3}$ 
for $\alpha=1.5$ (top-heavy).
Therefore, if only these massive stars become P-rich ONe novae during their binary
evolution, 
the required $N_{\rm ONe,0} \approx 0.01$ at [Fe/H]$=-3$
for $M_{\rm ej}=4 \times 10^{-5}$
can be marginally consistent with
top-heavy IMFs. 
This  required top-heavy IMFs 
is consistent with observational evidence of  more  top-heavy IMFs for
lower metallicities discovered  by Marks et al. (2012).

\begin{figure}
\epsscale{1.3}
\plotone{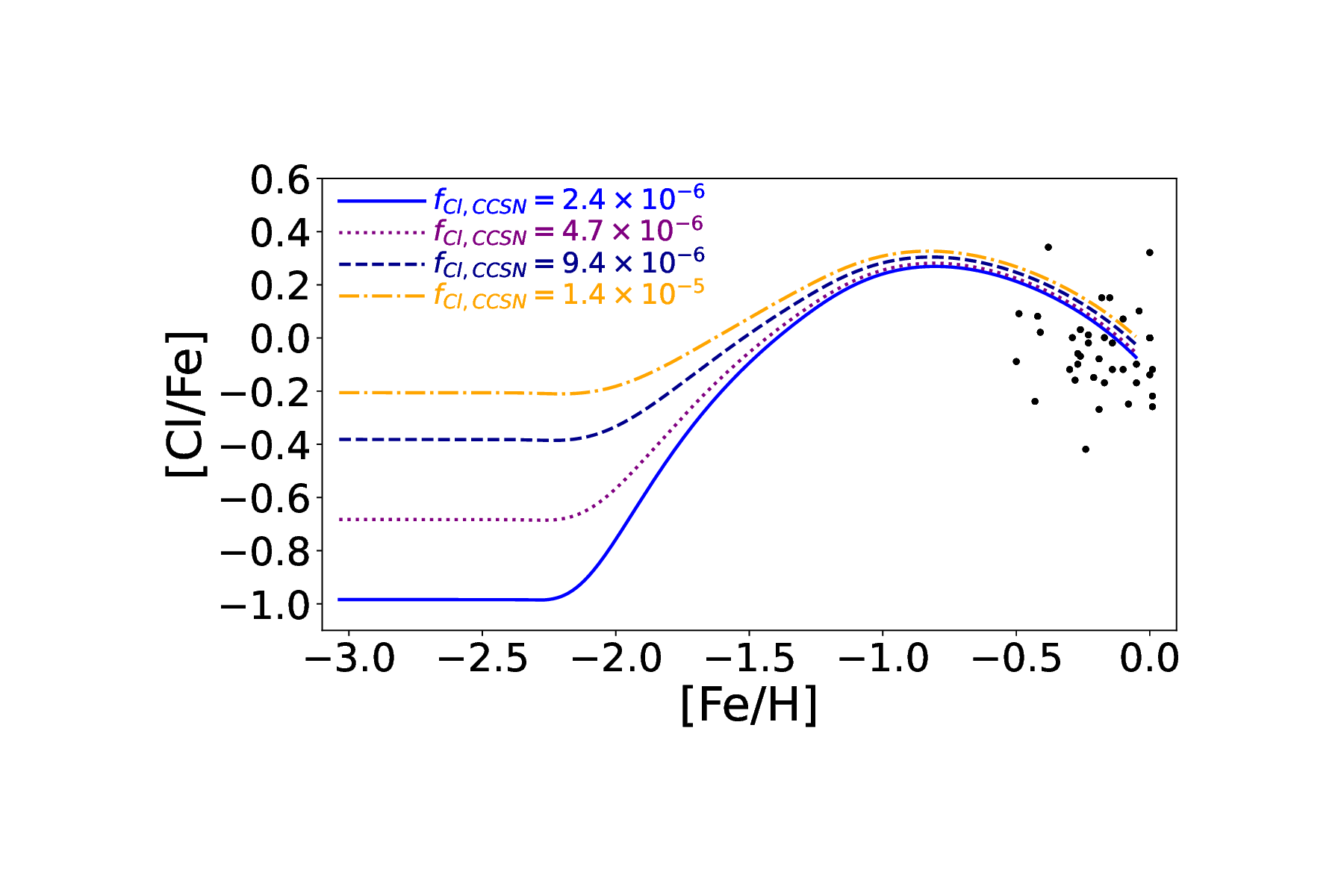}
\figcaption{
Evolutionary paths of four models with 
$f_{\rm Cl,CCSN}=2.4 \times 10^{-6}$ (blue solid),
$4.7 \times 10^{-6}$ (purple dotted),
$9.4 \times 10^{-6}$ (dark blue dashed),
and $1.4 \times 10^{-5}$ (orange dot-dashed)
on the [Fe/H]$-$[Cl/Fe] plane.
In these models,
all other parameter values are exactly the same as those in the fiducial model.
Observational data points for stars with $T_{\rm eff}>3500$K from 
Maas \& Pilachowski (2021)
are plotted by small black dots.
\label{fig-5}}
\end{figure}

ONe nova models with higher P yields 
($f_{\rm P}>10^{-3}$)
show lower $M_{\rm ej}$ ($<10^{-5} M_{\odot}$)
in previous studies (e.g., JH98;  S98;  Starrfield et al. 2024):
$M_{\rm ej}$ is only $5.2 \times 10^{-6} M_{\odot}$
for the model 8 with $f_{\rm p}=0.02$ in S98. 
Jos\'e et al. (2007) showed that
novae  with initial metallicities ($Z$) of
$10^{-7}$ and $2 \times 10^{-6}$ have $M_{\rm ej}=1.33 \times 10^{-5} M_{\odot}$
and $1.01 \times 10^{-5} M_{\odot}$, respectively, yet very high P yields
corresponding roughly to $\approx 500$ times the solar abundance. 
Although this implies that there are no previous novae models that 
reproduce both $f_{\rm P} \approx 0.01$ and
high $M_{\rm ej} \approx 4 \times 10^{-5} M_{\odot}$,
low-metallicity nova models with $10^{-6} < Z < 10^{-2}$ have not been explored yet.
The validity of the ONe nova scenario therefore depends crucially on whether these 
low-metallicity nova models
can predict both high $f_{\rm P}$ and high $M_{\rm ej}$.

The observed 
ejecta mass from V1974 Cyg 1992 
that is a candidate of 
ONe nova with $M_{\rm WD}=1.25 M_{\odot}$
ranges from $5 \times 10^{-5} M_{\odot}$ to 
$5 \times 10^{-4} M_{\odot}$ (e.g., Shore et al. 1993; Austin et al. 1996), which
is [10-100] times larger than the predicted values by S98.
If most  P-rich ONe novae can eject a larger amount
($\approx 10^{-4} M_{\odot}$) of gas like V1974 Cygni, 
the required $N_{\rm ONe,0}$ 
can be significantly lower for 
a reasonable  range of IMFs.
The observed large $M_{\rm ej}$ in V1974 Cygni, which is yet to be reproduced by existing nova models,
would demand the update of the ONe nova models too.

The present study has provided the following two key predictions that can be tested against
observations: (i)  [P/Fe] increases with increasing [Fe/H] 
for $-2 \lesssim {\rm [Fe/H]} \lesssim -1$ and (ii) the evolutionary path of the Galaxy
on the [Fe/H]$-$[Cl/Fe] plane
is very similar to that on the [Fe/H]$-$[P/Fe] plane.
It is therefore doubtlessly worthwhile for future observations 
to investigate
[P/Fe] of the Galactic halo and thick disk stars with 
for $-2 \lesssim  {\rm [Fe/H]} \lesssim -1$.
Such observations will also be able to distinguish between
chemical enrichment by ONe novae and by massive stars with O-C shell merging
(Ritter et al. 2018).
Likewise,  spectroscopic studies of [Cl/Fe] of stars with [Fe/H]$<-0.5$ are crucial
to assess the validity of the ONe nova scenario, though such observations could be 
challenging for existing telescopes.

M20 have discovered P-rich stars with unusually high [P/Fe] ($>1$)
and suggested that super-AGB stars and novae can produce a large amount of P
required to explain [P/Fe] of the P-rich stars.
Although the present study has so far focused on the [P/Fe]$-$[Fe/H] relation of P-normal
stars in the Galaxy, it can provide a possible solution to the origin
of P-rich stars discovered by
M20.
If P-rich ejecta from P-rich ONe novae is mixed with a smaller amount of ISM
without being much polluted by CCSNe and SNe Ia, new stars formed from the mixed gas
can have rather high [P/Fe].  A key question of this scenario is whether 
the observed abundances of  elements other than P  such as [O/Fe], [Mg/Fe], [Al/Fe],
and [Si/Fe]  can be
self-consistently explained by the scenario. Since JH98 predicted higher levels of
enhancement in the above elements, it is our future study to investigate
whether the observed chemical abundance patters of these elements in P-rich stars can be 
reproduced well by our future models. 
The observed higher [Ce/Fe] of P-rich stars
also needs to be investigated by our
future models incorporating chemical enrichment by super-AGB stars that are progenitor
of WDs for ONe novae and can eject Ce-rich winds (e.g., Doherty et al. 2014).

\section{Acknowledgment}
We are   grateful to the referee  for  constructive and
useful comments that improved this paper.
TT was supported by JSPS KAKENHI Grant Numbers 18H01258, 19H05811, and 23H00132.

{}


\begin{thebibliography}{}

\bibitem[]{}
Austin, S. J., et al. 1996, AJ, 111, 869

\bibitem[]{}
Bekki, K.,  \& Tsujimoto, T. 2012, ApJ, 761, 180

\bibitem[]{}
Bekki, K.,  \& Tsujimoto, T. 2023, MNRAS, 526, L26

\bibitem[]{}
Caffau, E., Bonifacio, P., Faraggiana, R., Steffen, M. 
2011, A\&A, 532, 98

\bibitem[]{}
Cescutti, G., et al. 2012, A\&A,  540, 33 (C12)

\bibitem[]{}
Cescutti, G., \&  Molaro, P. 2019, MNRAS, 482, 4372


\bibitem[]{}
Della Valle, M., \&  Izzo, L. 2020, A\&ARv, 28, 3

\bibitem[]{}
Doherty, C. L., et al. 2014, MNRAS, 441, 582


\bibitem[]{}
Hinkel, N. R., Hartnett, H. E.,  Young, P. A. 2020, ApJL, 900, 38

\bibitem[]{}
Jos\'e, J., \&  Hernanz, M. 1998, ApJ, 494, 680 (JH98)

\bibitem[]{}
Jos\'e, J., et al. 2007,  ApJL, 662, 103

\bibitem[]{}
Kemp, A. J., et al. 2022, MNRAS, 509, 1175 (K22)

\bibitem[]{}
Kobayashi, C., et al. 2006, ApJ, 653, 1145 (K06)

\bibitem[]{}
Maas, Z. G., \&  Pilachowski, C. A.  2021, AJ, 161, 183

\bibitem[]{}
Maas, Z. G., et al. 2022, ApJ, 164, 61
 
\bibitem[]{}
Marks, M., et al. 2012, MNRAS, 422, 2246

\bibitem[]{}
Masseron, T., et al. 2020, NatCo, 11, 3759 (M20)

\bibitem[]{}
Matteucci, F., et al. 2003, A\&A, 405, 23


\bibitem[]{}
Nandakumar, G., et al. 2022, A\&A, 668, 88

\bibitem[]{}
Prantzos, N., et al. 2018, MNRAS, 476, 3432


\bibitem[]{}
Ritter, C., et al. 2018, MNRAS, 474 L1
\bibitem[]{}
Roederer, I. U., et al. 2014, ApJ, 797, 69

\bibitem[]{}
Roederer, I. U., et al. 2016, ApJL, 824, 19

\bibitem[]{}
Romano, D.,  Matteucci, F. 2003, MNRAS, 342, 185

\bibitem[]{}
Shore, S.  N., et al. 1993, AJ, 106, 2408


\bibitem[]{}
Sneden, C., et al. 2021, AJ, 161, 128

\bibitem[]{}
Starrfield, S., et al. 1998, MNRAS, 296, 502 (S98)

\bibitem[]{}
Starrfield, S., et al. 2024, preprint (arXiv240102307)



\bibitem[]{}
Tremblay, P., et al. 2024, preprint (arXiv:2402.14960)

\bibitem[]{}
Truran, J. W., \&  Livio, M. 1986,  ApJ, 308, 721

\end{thebibliography}
\end{document}